\documentclass{iopart}

\usepackage{graphicx}
\usepackage{xcolor}
\usepackage{iopams}
\usepackage{cite}

\begin{document}
\title[Disordered exclusion process in the low-current regime]{Disordered exclusion process revisited: some exact results in the low-current regime}
\author{J Szavits Nossan}
\address{$^1$ SUPA, School of Physics and Astronomy, University of Edinburgh, Mayfield Road, Edinburgh EH9 3JZ, United Kingdom}
\address{$^2$ Institute of Physics, Bijeni\v{c}ka cesta 46, HR-10001 Zagreb, Croatia}
\ead{jszavits@staffmail.ed.ac.uk}

\begin{abstract}
We study steady state of the totally asym\-me\-tric sim\-ple ex\-clu\-sion pro\-cess  with in\-ho\-mo\-geneous hopping rates associated with sites (site-wise disorder). Using the fact that the non-normalized steady-state weights which solve the master equation are polynomials in all the hopping rates, we propose a general method for calculating their first few lowest coefficients exactly. In case of binary disorder where all slow sites share the same hopping rate $r<1$, we apply this method to calculate steady-state current up to the quadratic term in $r$ for some particular disorder configurations. For the most general (non-binary) disorder, we show that in the low-current regime the current is determined solely by the current-minimizing subset of equal hopping rates, regardless of other hopping rates. Our approach can be readily applied to any other driven diffusive system with unidirectional hopping if one can identify a hopping rate such that the current vanishes when this rate is set to zero.
\end{abstract}
\submitto{\JPA}

\maketitle


\section{Introduction}

Despite numerous efforts conducted in the past, the understanding of macroscopic systems out of equilibrium is still far from being put in a systematic theory. Unlike the relaxation around the equilibrium which has become a standard textbook material, less can be said for systems that are maintained far from equilibrium. One possible route to fill this gap that proved useful in the past (e.g. in developing general theory of critical phenomena) is to study particular microscopic models. Usually, one starts with a minimal model that admits analytical treatment (exact or approximate), later adding more details to make it more realistic. Unfortunately, the lack of detailed balance - a defining property of systems maintained far from equilibrium - means that we are generally deprived even of the knowledge of its steady state, not to mention the relaxation mechanism towards it. Exactly solved models far from equilibrium are thus rare, but have an important role in nonequilibrium statistical mechanics.

One such model is the asymmetric simple exclusion process (ASEP), a minimal model of transport of (classical) particles driven by an external field and interacting only through the exclusion principle that prevents them from coming too close to each other. Although simplified, this interaction describes
several real situations on various length scales, ranging from mobile ions in superionic conductors \cite{MarroDickman99} to self-propelled particles in mesoscopic (ribosomes on a mRNA \cite{MacDonaldGibbsPipkin68,MacDonaldGibbs69}) and macroscopic (cars \cite{NagelSchreckenberg92}) systems. From a theoretical viewpoint, the ASEP has become a paradigmatic model of boundary-induced phase transitions that are, unlike their equilibrium counterpart (with short-range interactions), present even in one dimension. Originally proposed to model translation of mRNA more than five decades ago \cite{MacDonaldGibbsPipkin68,MacDonaldGibbs69}, the exact solution in 1993 \cite{SchutzDomany93,DEHP} sparked a great interest that led to  several important results relevant to the general theory of nonequilibrium steady states. Using the exact solution of the ASEP, Derrida, Lebowitz and Speer \cite{DerridaLebowitzSpeer01} derived a non-equilibrium analogue of the free energy that describes coarse-grained fluctuations around the steady state, thus generalizing older Onsager-Machlup theory \cite{OnsagerMachlup53} valid around equilibrium. This later helped Bertini \etal \cite{Bertini09} to build the macroscopic theory of fluctuations for driven diffusive systems, of which the ASEP is just one example.

A lot of work has been devoted to improve the ASEP to better fit particular phenomena, most of which have their origins in vehicular traffic or biology (for a comprehensive and recent review see \cite{Schadschneider10}). The exact solution has been successfully extended to particle-dependent hopping rates \cite{Mallick96,LeePopkovKim97,Evans96,KrugFerrari96} and multispecies systems \cite{MallickMallickRajewsky99,EvansFerrariMallick09}, both of importance for traffic phenomena. In biology, some of the generalizations include particles occupying more than one site \cite{LakatosChou03,ShawZiaLee03}, position-dependent hopping rates (termed site-wise disorder) \cite{ChouLakatos04}, desorption and adsorption of particles in the bulk (i.e. Langmuir kinetics) \cite{Parmeggiani03}, particles with internal states \cite{Reichenbach06}, extension to more than one lane \cite{EvansKafriSugdenTailleur11}, dynamically extending lattice \cite{SugdenEvans07}, etc. Unfortunately, common to most of these generalizations is non-applicability of the exact methods that were used to solve the original ASEP, namely the matrix-product ansatz (see e.g. \cite{BlytheEvans07}). Instead, most approaches utilize various mean-field approximations that neglect correlations in a fashion similar to truncation in the Bogoliubov-Born-Green-Kirkwood-Yvon (BBGKY) hierarchy. While they give a satisfactory account for most of the aforementioned phenomena, it is less clear how to improve them in some controlled fashion. In some cases such as site-wise disorder, the mean-field approximation itself becomes hard to treat analytically as the number of inhomogeneities increases. 

In this article we study totally asymmetric simple exclusion process (TASEP) - a version of the ASEP with unidirectional hopping - in the presence of site-wise disorder and show that some of the information about its exact steady state can still be retained. Our starting point is the known fact, reviewed in \ref{appendix_a}, that the non-normalized steady-state weights which solve the master equation can always be written as multivariate polynomials in all the hopping rates or alternatively, as univariate polynomials in one of the hopping rates with polynomial coefficients that depend on all other hopping rates \cite{EvansBlythe02}. As our main result, we show how to compute the first few lowest polynomial coefficients exactly, which gives a good approximation of the exact steady-state weights if the hopping rate that we expand in is much smaller than the other rates. Since in that case the particle current is small, we term this regime the low-current regime. Our method of computing non-normalized steady-state weights in the low-current regime is not restricted to site-wise disorder but can be readily applied to many other generalizations of the TASEP and other driven diffusive system with unidirectional hopping.

The paper is organized as follows. In section \ref{model} we define the TASEP with site-dependent hopping rates and review some of the related exact results. A general approach for calculating steady state in the low-current regime is devised in section \ref{general} for arbitrary disorder distributions. Section \ref{examples} is devoted to few particular cases of binary disorder where all the slow sites share the same hopping rate $r<1$. Our main interest there is to calculate steady-state average of the particle current, i.e. its Taylor expansion in one of the hopping rates which is considered to be small. Some of the exact results that we obtain have already been known, but here they are derived for the first time rather then being guessed from the exact solution of small systems \cite{JanowskyLebowitz94,LakatosChou03}. Further applications, mainly devoted to the most general case of non-binary disorder, are discussed in \ref{further_applications} with some interesting new implications for the protein synthesis.


\section{Model}
\label{model}

\begin{figure}[bht]
\centering\includegraphics[width=4cm]{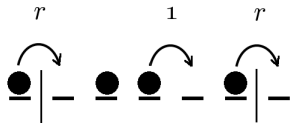}
\caption{Schematic picture of the TASEP with binary disorder. Bonds where particles jump at rate $r<1$ are represented by (permeable) walls to emphasize that letting $r=0$ means no particle is allowed to move through the wall.}
\label{fig1}
\end{figure}

We consider the totally asymmetric simple exclusion process (TASEP) on an one-dimensional lattice of $L$ sites, where each site $i=1,\dots, L$ is either occupied by a particle ($\tau_i=1$) or empty ($\tau_i=0$). Particles each move forward stochastically at rate $p_i$ (which is modelled by the random-sequential update), provided the site $i+1$ in front is empty (exclusion principle). In this paper we consider mainly binary disorder \cite{TripathyBarma97,TripathyBarma98}, which means that $p_i$ is either $r<1$ (slow sites) or $1$ (regular sites). (Non-binary disorder with $p_i\in\mathbf{R}$ to which our approach applies as well is left for section \ref{further_applications}.) Boundary conditions can be either periodic ($\tau_{L+1}=\tau_1$) or open, the latter meaning that new particles are injected at site site $i=1$ at rate $\alpha$ provided it is empty, and are removed from the site $i=L$ at rate $\beta$. An illustration of the process (not drawing boundary rules explicitly) is presented in figure \ref{fig1}.


The steady state is described completely by the probability $P(C)$ to find the system in a given configuration $C=\{\tau_1,\dots,\tau_L\}$, which solves the following continuous-time master equation, written here in a compact form,

\numparts
\begin{eqnarray}
\label{master_periodic}
\mathrm{(periodic)}\quad &0&=\sum_{i=1}^L p_i(\tau_i-\tau_{i+1})P(\dots,\tau_{i-1},1,0,\tau_{i+2},\dots)\\
\label{master_open}
\mathrm{(open)}\quad &0&=-\alpha(1-2\tau_1)P(0,\dots)+\beta(1-2\tau_L)P(\dots,1)+\nonumber\\
&&\qquad+\sum_{i=1}^{L-1} p_i(\tau_i-\tau_{i+1})P(\dots,\tau_{i-1},1,0,\tau_{i+2},\dots).
\end{eqnarray}
\endnumparts

In \ref{appendix_a} we show that the steady-state probability $P(C)$ is proportional to the determinant of a matrix whose matrix elements are linear combinations of hopping rates. That allows us to write $P(C)$ in the following form,

\begin{equation}
\label{P}
P(C)=\frac{f(C)}{\sum_{C}f(C)},
\end{equation}

\noindent where $f(C)$ is a \emph{multivariate polynomial} in all hopping rates,

\numparts
\begin{eqnarray}
\label{poly_periodic}
\textrm{(periodic)}\quad && f(C)=\sum_k f_{k}(C)r^k\\
\label{poly_open}
\textrm{(open)}\quad && f(C)=\sum_{i,j,k} f_{i,j,k}(C)\alpha^{i}\beta^{j}r^{k}.
\end{eqnarray}
\endnumparts

\noindent The particular form of (\ref{P}) with $f(C)$ given by (\ref{poly_periodic}) or (\ref{poly_open}) means that the ensemble average of any physical observable $g(C)$ (e.g. local density $\tau_i$ at site $i$, or current $p_i\tau_i(1-\tau_{i+1})$) is a rational function with respect to any of the hopping rates (say $r$)

\begin{equation}
\langle g(C)\rangle=\sum_{C}g(C)P(C)=\frac{\sum_C g(C)f(C)}{\sum_C f(C)}=\frac{\sum_{k}a_k r^k}{\sum_{k}b_k r^k},
\label{rational}
\end{equation}

\noindent where coefficients $a_k$ and $b_k$ are given by

\begin{equation*}
a_k=\sum_C g(C)f_k(C), \quad b_k=\sum_C f_k(C).
\end{equation*}

\noindent By expanding (\ref{rational}) in Taylor series around $r=0$ we arrive at 

\begin{equation}
\langle g(C)\rangle=\sum_{k=0}^{\infty}c_k r^k,
\label{expansion_g}
\end{equation}

\noindent where

\begin{equation*}
c_k=\cases{\frac{a_k}{b_0}-\sum_{n=1}^{k}c_{k-n}\frac{b_n}{b_0}, & $k=0,1,\dots,\mathcal{M}$\cr
-\sum_{n=1}^{\mathcal{M}}c_{k-n}\frac{b_n}{b_0}, & $k\geq\mathcal{M}+1$.\cr}
\end{equation*}

\noindent Here $\mathcal{M}$ is the maximal degree of all $f(C)$'s as polynomials in $r$, and can be calculated by invoking Schnakenberg's network theory \cite{Schnakenberg1976}. The value of $\mathcal{M}$ is not important in our approach, as calculating coefficients beyond first few terms in (\ref{expansion_g}) becomes extremely difficult.

The idea of expanding $\langle g(C)\rangle$ in small $r$ originates from the work of Janowsky and Lebowitz \cite{JanowskyLebowitz94}. They obtained the first few coefficients in the expansion of current $j_L(r)$ around $r=0$ in the TASEP with a single slow site both in the periodic and in the open boundaries case with $r\ll\alpha=\beta$

\begin{equation}
j_L(r)=r-\frac{3}{2}r^2+\frac{19}{16}r^3-\frac{21535}{27648}r^4+\frac{77729356627}{146767085568}r^5-\Or(r^6).
\label{j_1}
\end{equation}

\noindent The expansion (\ref{j_1}) was obtained by solving the full master equation for small systems with $L\leq 8$ and noticing that as $L$ increases, low-order terms become independent of $L$. A similar approach, coined \emph{finite segment mean-field theory} (FSMFT), has been devised by Chou and Lakatos in \cite{ChouLakatos04} for the open boundaries case with few slow sites. When slow sites are confined to a small segment, the idea is to solve the master equation exactly for that small segment and then calculate the current $j(\sigma_-,\sigma_+)$ as a function of the average densities $\sigma_-$ and $\sigma_+$ at its ends. Assuming that the density profiles are flat outside the segment, $\sigma_-$ and $\sigma_+$ can be then calculated numerically by solving $\sigma_-(1-\sigma_-)=j(\sigma_-,\sigma_+)=\sigma_+(1-\sigma_+)$. For $r\ll\alpha,\beta>1/2$, Chou and Lakatos were able to deduce the analytical expression for the current up to $\Or(r^2)$ for two particular configurations of disorder, namely

\begin{equation}
j_L(r)=\frac{d}{d+1}r+\Or(r^2),
\label{j_2}
\end{equation}

\noindent for two slow sites placed $d$ sites apart, and

\begin{equation}
\label{j_bottle}
j_L(r)=\frac{l+1}{4l-2}r+\Or(r^2),
\end{equation}

\noindent for a bottleneck of $l$ slow sites. As $d\rightarrow\infty$ in (\ref{j_2}), the current for $r\ll \alpha,\beta>1/2$ approaches $j_L(r)\approx r+\Or(r^2)$ as in the TASEP with one slow site. In (\ref{j_bottle}), the current is clearly dominated by the capacity of the bottleneck and approaches $r/4$ when $l\rightarrow\infty$. The downside of FSMFT is that the present computing power restricts the size of the segment that is treated exactly to $\approx 20$ sites, thus excluding more complex disorder configurations. Also, FSMFT is basically a brute force attack on the master equation and tells us little about where the coefficients e.g. in (\ref{j_1})-(\ref{j_bottle}) come from. In the next section we will present a general approach for computing low-order terms in the expansion $f(C)=f_0(C)+\gamma f_1(C)+\Or(\gamma^2)$, where $\gamma$ stands for any of the model's hopping rates such that the current $j_L(\gamma)\rightarrow 0$ as $\gamma\rightarrow 0 $ ($\gamma=r$ in the periodic boundaries case or $\gamma\in\{r,\alpha,\beta\}$ in the open boundaries case). 


\section{Main idea and general results}
\label{general}

We start by writing $f(C)$ as a polynomial in one of the model's hopping rates $\gamma$,

\begin{equation}
f(C)=\sum_k f_k(C)\gamma^k,
\label{gamma}
\end{equation}

\noindent where we assume $f_k(C)$ to be dependent on $C$ and all other hopping rates $\neq\gamma$. Inserting (\ref{gamma}) in the stationary master equation (\ref{master_periodic}) or (\ref{master_open}) and collecting all the terms of $\Or(\gamma^k)$ gives recursion relations

\begin{equation}
\label{master_f0}
\eqalign{0&=\sum_{C'}(1-\delta_{\gamma,W(C'\rightarrow C)})W(C'\rightarrow C)f_{0}(C')\cr
&-\sum_{C'}(1-\delta_{\gamma,W(C\rightarrow C')})W(C\rightarrow C')f_{0}(C)\cr}
\end{equation} 

\begin{equation}
\fl\eqalign{0&=\sum_{C'}(1-\delta_{\gamma,W(C'\rightarrow C)})W(C'\rightarrow C)f_{k}(C')+\sum_{C'}\delta_{\gamma,W(C'\rightarrow C)})f_{k-1}(C')\cr
\fl &-\sum_{C'}(1-\delta_{\gamma,W(C\rightarrow C')})W(C\rightarrow C')f_{k}(C)-\sum_{C'}\delta_{\gamma,W(C\rightarrow C')})f_{k-1}(C),\quad k>0.\cr}
\label{master_fk}
\end{equation}

\noindent It is useful to picture our system as a directed graph made of vertices (configurations) and directed edges (transitions between configurations) weighted by the hopping rates. Edges that are weighted by $\gamma$ are called slow, and all the others are called regular. A special role here is played by configurations that have all their outgoing edges slow. We will call such configurations \emph{blocked} (with respect to $\gamma$) because if the system gets into one of these configurations and $\gamma\rightarrow 0$, it will stay there forever. If $B$ is a non-empty set of all such configurations then terms $f_0(C)$ are clearly absent from (\ref{master_f0}) for any $C\in B$. For example, the TASEP with a slow site and periodic boundary conditions has only one blocked configuration with respect to $r$, the one in which all particles are behind the slow site. In the open boundaries case there are two additional blocked configurations, one with respect to $\alpha$ (an empty lattice) and one with respect to $\beta$ (full lattice). 

The fact that some terms are missing when collecting zeroth-order terms may seem confusing, because we have to go to the next order in $\gamma$ to calculate $f_0(C)$, but we need $f_0(C)$ to calculate the next order terms. In the next section we show how to eliminate all first-order terms, thus ending up with a closed set equations for $f_0(C)$, $C\in B$.

\subsection{Zeroth-order terms}

We start by outlining the general procedure for calculating zeroth-order terms and then show how it works on an explicit example. Since $f_0(C)$ is absent from (\ref{master_f0}) for any $C\in B$, we can assume that $f_0(C)\neq 0$ for $C\in B$ and then solve (\ref{master_f0}) by setting all the remaining zeroth-order terms to $0$,

\begin{equation}
f_0(C)=0,\quad C\notin B.
\label{f0_regular}
\end{equation}

\noindent Thus the existence of blocked configurations reduces the calculation of $f_0(C)$ to a smaller set of configurations $B$. 

To get a closed set of equations for the remaining zeroth-order terms we have to inspect terms that are linear in $\gamma$,

\begin{equation}
\fl \eqalign{0=&\sum_{C'}(1-\delta_{\gamma,W(C'\rightarrow C)})W(C'\rightarrow C)f_{1}(C')+\sum_{C'}\delta_{\gamma,W(C'\rightarrow C)})f_{0}(C')\cr
\fl &-\sum_{C'}(1-\delta_{\gamma,W(C\rightarrow C')})W(C\rightarrow C')f_{1}(C)-\sum_{C'}\delta_{\gamma,W(C\rightarrow C')})f_{0}(C).\cr}
\label{master_f1}
\end{equation}

\noindent It will prove useful to write (\ref{master_f1}) in the following form,

\begin{equation}
f_1(C)=\sum_{C'\in B}\lambda(C,C')f_0(C'), \quad C\notin B,
\label{recursion_1}
\end{equation}

\noindent where $\lambda(C,C')$ is some unknown matrix. Inserting (\ref{recursion_1}) in (\ref{master_f1}), the equation for $f_0(C)$ now reads

\begin{equation}
\mathcal{A}_\gamma(C)f_0(C)=\sum_{C''\in B}\kappa(C,C'')f_0(C''),\quad C\in B
\label{system_f0}
\end{equation}

\noindent where $\mathcal{A}_\gamma(C)$ and $\kappa(C,C'')$ are given by

\begin{equation}
\mathcal{A}_\gamma(C)=\sum_{C'}\delta_{\gamma,W(C\rightarrow C')},
\label{A_gamma}
\end{equation}

\begin{equation}
\kappa(C,C'')=\sum_{C'}(1-\delta_{\gamma,W(C'\rightarrow C)})\lambda(C',C'')+\delta_{\gamma,W(C''\rightarrow C)}.
\label{kappa}
\end{equation}

\noindent To find $f_0(C)$, we have to find $\lambda(C,C')$, calculate $\kappa(C',C'')$ and then solve (\ref{system_f0}). Luckily, $\kappa(C',C'')$ can be found without using the expression (\ref{kappa}). The algorithm that we give below is essentially the same as the one we'll use later for constructing $\lambda(C,C')$.

To start with, let's call a path $\mathcal{P}$ in configuration space any sequence of configurations $C_1,C_2,\dots,C_n$ such that none of $W(C_1\rightarrow C_2)$, $W(C_2\rightarrow C_3)$, $\dots$, $W(C_{n-1}\rightarrow C_n)$ are zero. A regular path is a path in which none of the edges is slow. With all these preliminaries, we rewrite the equation (\ref{master_f1}) for $f_0(C)$, $C\in B$,

\begin{equation}
\eqalign{\mathcal{A}_\gamma(C)f_0(C)&=\sum_{C'}W(C'\rightarrow C)(1-\delta_{\gamma,W(C'\rightarrow C)})f_{1}(C')\cr
&+\sum_{C'}\delta_{\gamma,W(C'\rightarrow C)}f_{0}(C'),\quad C\in B.\cr}
\label{master_f1a}
\end{equation}

\noindent Configurations $C'$ on the r.h.s. are obtained by moving one of the particles \emph{backwards}. Any move to a configuration $C'\notin B$ across a slow edge should be discarded since $f_0(C')=0$ for $C'\notin B$ and the second term on the r.h.s. vanishes in that case. Now let's focus on $C'$ for which $W(C'\rightarrow C)\neq\gamma$. Since we started from $C\in B$ and moved one particle backwards across a regular edge, $C'$ has the following properties: (a) $C'\notin B$, (b) $f_0(C')=0$ and (c) $\mathcal{A}_0(C')=1$. The last one is simply because $C'$ is just one jump from $C$ and therefore cannot have any other regular outgoing edges except the one pointing towards $C$. Now, let's write the equation for $f_1(C')$,

\begin{equation}
\eqalign{\underbrace{\mathcal{A}_0(C')}_{=1}\cdot f_1(C')&+\mathcal{A}_\gamma(C')\underbrace{f_0(C')}_{=0}=\sum_{C''}\delta_{\gamma,W(C''\rightarrow C')}f_{0}(C'')\cr
&+\sum_{C''}W(C''\rightarrow C')(1-\delta_{\gamma,W(C''\rightarrow C')})f_{1}(C'').\cr}
\label{master_f1b}
\end{equation}

\noindent The l.h.s. of (\ref{master_f1b}) is exactly what we get on the r.h.s. of (\ref{master_f1a}) by moving one particle across a regular edge from $C\in B$. We can then insert (\ref{master_f1b}) in (\ref{master_f1a}) and thus eliminate $f_1(C')$. The idea is to repeat this process of moving particles backwards and substituting $f_1(C'')$ from 

\begin{equation}
\eqalign{\mathcal{A}_0(C'')f_1(C'')&+\mathcal{A}_\gamma(C'')\underbrace{f_0(C')}_{=0}=\sum_{C'''}\delta_{\gamma,W(C'''\rightarrow C'')}f_{0}(C''')\cr
&+\sum_{C'''}W(C'''\rightarrow C'')(1-\delta_{\gamma,W(C'''\rightarrow C'')})f_{1}(C'''),\cr}
\label{master_f1c}
\end{equation}

\noindent for any $C''\notin B$ that we reach along. Mathematically speaking, we must exhaust all backward paths originating from $C$ that have one slow edge at the end and all other edges regular. Moreover, since $f_0(C')=0$ for all $C'\notin B$, we should consider only paths that end in configurations belonging to $B$. 

Now, consider any $C''\notin B$ that is reached by moving particles backwards from $C$ without crossing a slow edge. Then if we start at $C''$ and move particles forward without crossing a slow edge, we must end at $C$. This ensures that by moving particles backwards in order to eliminate any $f_1(C'')$ along the way, $f_1(C'')$ will appear \emph{exactly} $\mathcal{A}_0(C'')$ times. We can then complete the l.h.s. of (\ref{master_f1c}) (since $\mathcal{A}_\gamma(C'')f_0(C'')=0$) and substitute (\ref{master_f1c}) in (\ref{master_f1a}). In the end, when all paths have been exhausted and all first-order terms eliminated, the final result is

\begin{equation}
\mathcal{A}_\gamma(C)f_0(C)=\sum_{C'\in S_0(C)}f_0(C'), \quad C\in B
\label{system_f0a}
\end{equation}

\noindent where $S_0(C)$ is the set of all configurations $C'$ such that (a) $C'$ can be reached from $C$ by moving particles backwards and crossing a slow edge in the last move only and (b) $f_0(C')\neq 0$. (From this definition, $S_0(C)\subset B$ for any $C\in B$.) Going back to (\ref{kappa}) we have therefore proved that

\begin{equation*}
\kappa(C,C')=\cases{1, & $C\in B$ and $C'\in S_0(C)$,\cr
		    0, & otherwise.\cr}
\end{equation*}


\begin{figure}[bht]
\centering\includegraphics[width=7cm]{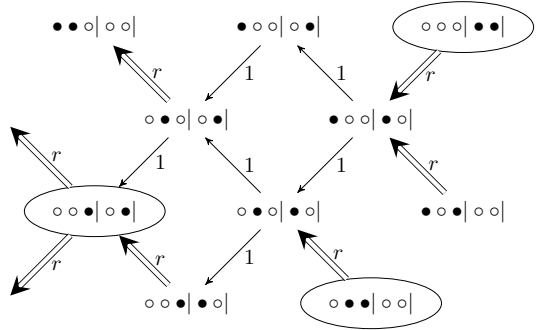}
\caption{A part of the graph representing the TASEP on a ring of $L=5$ sites with $N=2$ particles and slow sites placed at $i=2$ and $i=5$. Blocked configurations are designated with ellipses and slow edges with double arrows.}
\label{fig2}
\end{figure}

To illustrate this, let's consider the TASEP on a ring with $L=5$ sites, $N=2$ particles and two slow sites placed at $i=3$ and $i=5$. There are $5!/(2!3!)=30$ configurations in total of which $\vert B\vert=3$ are blocked, $B=\{\circ\bullet\bullet\vert\circ\circ\vert,\,\circ\circ\bullet\vert\circ\bullet\vert,\,\circ\circ\circ\vert\bullet\bullet\vert\}$ (as in figure \ref{fig1}, sites in front of the walls are considered slow). A portion of the graph containing configuration $C=\circ\circ\bullet\vert\circ\bullet\vert$ is presented in figure \ref{fig2}. The equation for $f_0(\circ\circ\bullet\vert\circ\bullet\vert)$ reads

\begin{equation}
\label{example}
\fl \eqalign{2f_0(\circ\circ\bullet\vert\circ\bullet\vert)&=f_1(\circ\bullet\circ\vert\circ\bullet\vert)+f_1(\circ\circ\bullet\vert\bullet\circ\vert)=\cr
\fl &=[f_1(\circ\bullet\circ\vert\circ\bullet\vert)+\underbrace{f_0(\circ\bullet\circ\vert\circ\bullet\vert)}_{=0}]+f_1(\circ\circ\bullet\vert\bullet\circ\vert),\cr}
\end{equation}

\noindent where in the second line we added $f_0(\circ\bullet\circ\vert\circ\bullet\vert)=0$ to complete the master equation for $f_1(\circ\bullet\circ\vert\circ\bullet\vert)$,

\begin{equation}
\label{example1a}
f_1(\circ\bullet\circ\vert\circ\bullet\vert)+f_0(\circ\bullet\circ\vert\circ\bullet\vert)=f_1(\bullet\circ\circ\vert\circ\bullet\vert)+f_1(\circ\bullet\circ\vert\bullet\circ\vert).
\end{equation}

\noindent Master equation for the remaining term $f_1(\circ\circ\bullet\vert\bullet\circ\vert)$ reads

\begin{equation}
\label{example1b}
f_1(\circ\circ\bullet\vert\bullet\circ\vert)=f_1(\circ\bullet\circ\vert\bullet\circ\vert).
\end{equation}

\noindent Substituting (\ref{example1a}) and (\ref{example1b}) in (\ref{example}), the equation for $f_0(\circ\circ\bullet\vert\circ\bullet\vert)$ now reads

\begin{equation}
\label{example2}
2f_0(\circ\circ\bullet\vert\circ\bullet\vert)=f_1(\bullet\circ\circ\vert\circ\bullet\vert)+2f_1(\circ\bullet\circ\vert\bullet\circ\vert).
\end{equation}

\noindent Equations for the terms on the r.h.s. of (\ref{example2}) are

\begin{equation*}
f_1(\bullet\circ\circ\vert\circ\bullet\vert)=f_1(\bullet\circ\circ\vert\bullet\circ\vert),
\end{equation*}

\begin{equation*}
2f_1(\circ\bullet\circ\vert\bullet\circ\vert)=f_1(\bullet\circ\circ\vert\bullet\circ\vert)+f_0(\circ\bullet\bullet\vert\circ\circ\vert),
\end{equation*}

\noindent which upon substitution in (\ref{example2}) becomes

\begin{equation}
\label{example3}
2f_0(\circ\circ\bullet\vert\circ\bullet\vert)=2f_1(\bullet\circ\circ\vert\bullet\circ\vert)+f_0(\circ\bullet\bullet\vert\circ\circ\vert).
\end{equation}

\noindent Finally, inserting

\begin{equation*}
2f_1(\bullet\circ\circ\vert\bullet\circ\vert)=\underbrace{f_0(\bullet\circ\bullet\vert\circ\circ\vert)}_{=0}+f_0(\circ\circ\circ\vert\bullet\bullet\vert),
\end{equation*}
\noindent in (\ref{example3}) gives the final equation for $f_0(\circ\circ\bullet\vert\circ\bullet\vert)$

\begin{equation}
\label{example4}
2f_0(\circ\circ\bullet\vert\circ\bullet\vert)=f_0(\circ\bullet\bullet\vert\circ\circ\vert)+f_0(\circ\circ\circ\vert\bullet\bullet\vert).
\end{equation}

\noindent This daunting task that we have just performed in fact has a remarkably simple interpretation. To see it, let's write equations for the remaining blocked configurations,

\begin{eqnarray}
\label{2}
&& f_0(\circ\bullet\bullet\vert\circ\circ\vert)=f_0(\circ\circ\bullet\vert\circ\bullet\vert),\\
\label{3}
&& f_0(\circ\circ\circ\vert\bullet\bullet\vert)=f_0(\circ\circ\bullet\vert\circ\bullet\vert).
\end{eqnarray}

\noindent The process described by (\ref{example4})-(\ref{3}) can be interpreted as follows: any particle that jumps from a slow site immediately joins the queue in front, while at the same time the whole queue that the particle has just left moves one step forward. Intuitively, this is easy to understand: the limit $r\rightarrow 0$ creates a huge separation of time scales, so that any particle that jumps from a slow sites is likely to join the queue in front before any other jump from a slow sites happens. Particles are thus exchanged between compartments separated by walls, each compartment having a finite capacity which is the number of sites between two neighbouring walls. This number can be greater than $1$, as in our example above, and so the process that TASEP maps to in the limit $r\rightarrow 0$ is a generalization of the exclusion process called the \emph{partial} exclusion process (not to be confused with \emph{partially asymmetric} exclusion process). Interestingly, the partial exclusion process has been introduced long time ago by Sch\"{u}tz \cite{SandowSchutz94}, but has been rarely studied since \cite{Schutz03,Thompson11}.


\begin{figure}[bht]
\centering\includegraphics[width=6cm]{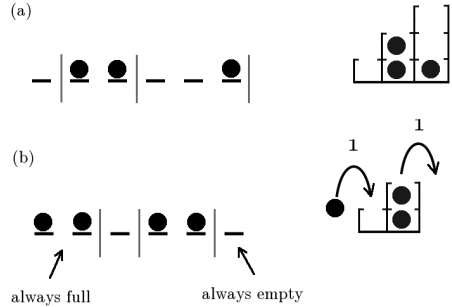}
\caption{Mapping the TASEP with binary disorder in the limit $r\rightarrow 0$ (left column) to a partial exclusion process (right column) in (a) periodic boundaries case and (b) open boundaries case. In both cases particles move to the right.}
\label{fig3}
\end{figure}

For zeroth-order terms, our result can be summarized as follows: consider the TASEP with periodic boundary conditions on a lattice of $L$ sites, $N$ particles and $D$ slow sites placed at sites $k_1,k_2,\dots,k_D=L$ (due to periodic boundary conditions, we have a freedom to place the last slow site at the end). Then $f_0(C)$ for $C\in B$ solves the master equation of the totally asymmetric \emph{partial} exclusion process on a lattice of $D$ sites, where each site $i=1,\dots,D$ can hold at most $k_{i}-k_{i-1}$ particles, where $k_0=1$ (see figure \ref{fig3}a). 

One striking thing about this result is that $\kappa(C,C')$ and therefore any $f_0(C)$ are completely independent of the other hopping rates (if they exist) and instead depend only on the ``connectivity'' of blocked configurations. We will go back to this observation more in section \ref{further_applications} where we discuss two or more different slow hopping rates. For now, let's just focus on what this means for the binary disorder in the open boundaries case. In the open boundaries case, all blocked configurations (with respect to $r$) have the first compartment occupied by particles and the last one empty. If we have a lattice of $L$ sites and $D$ slow sites placed at $k_1,\dots,k_D$, then $f_0(C)$ for $C\in B$ solves master equation of a totally asymmetric partial exclusion process on a lattice of $D-1$ sites, where each site $i=1,D-1$ holds at most $k_{i+1}-k_i$ particles. At the boundaries, particles jump into the first compartment at rate $1$ if it's not full and leave the lattice from the last compartment at rate $1$ (see figure \ref{fig3}b). Thus the rates at which particles are exchanged with reservoirs are always maximal, i.e. equal to $1$.

For a small number of blocked configurations we may hope to solve (\ref{system_f0a}) analytically (as in section \ref{examples}) or numerically (even for $L$ large). As the number of blocked configurations increases, since the exact solution for the steady state of the totally asymmetric partial exclusion process is not known, we may end up with a problem no less harder that the one we started with. In the next section we give a general recipe for how to calculate first-order terms $f_1(C)$ if we can somehow solve (\ref{system_f0a}).

\subsection{First-order terms}
\label{first-order}

To calculate $f_1(C)$, $C\notin B$, we have to determine $\lambda(C,C')$ in (\ref{recursion_1}). The equation for $f_1(C)$ for any $C\notin B$ reads

\begin{equation}
\eqalign{f_1(C)&=\frac{1}{\mathcal{A}_0(C)}\sum_{C''\in B}\delta_{\gamma,W(C''\rightarrow C)}f_{0}(C'')\cr
&+\sum_{C''}\frac{W(C''\rightarrow C)}{\mathcal{A}_0(C)}(1-\delta_{\gamma,W(C'\rightarrow C)})f_{1}(C''),\cr}
\label{lambda_f1}
\end{equation}

\noindent where in the first sum we have used the fact that $f_0(C'')=0$ for all $C''\notin B$. Leaving all zeroth-order terms intact, we insert expressions like (\ref{lambda_f1}) recursively for the remaining first-order terms, until we are left with zeroth-order terms only. Again, we end up with as many zeroth-order terms as there are blocked configurations that can be reached from $C$ by moving particles backwards but crossing a slow edge in the last move only. Recalling that the set of all such configurations is $S_0(C)$, let's index a backward path from $C$ to any $C'\in S_0(C)$ with $\mathcal{P}_{C,C'}=C,C_1,C_2,\dots,C_n,C'$ ($n$ can, of course, vary from path to path). For any given $C$, $\lambda(C,C')$ is then given by

\begin{equation}
\fl\lambda(C,C')=\sum_{\mathcal{P}_{C,C'}}\frac{W(C_1\rightarrow C)}{\mathcal{A}_0(C)}\frac{W(C_2\rightarrow C_1)}{\mathcal{A}_0(C_1)}\cdot\dots\cdot\frac{W(C_{n-2}\rightarrow C_{n-1})}{\mathcal{A}_0(C_{n-1})}\frac{1}{\mathcal{A}_0(C_n)},
\label{path_weights}
\end{equation}

\noindent for $C'\in S_0(C)$ and is $0$ for $C'\notin S_0(C)$. 

To calculate $f_1(C)$ for $C\in B$, we use the same recipe as for $f_0(C)$. Starting from $C\in B$, we look for paths $\mathcal{P}_{C,C'}$ from $C$ to any $C'$ such that (a) $C'$ can be reached from $C$ by moving particles backwards and crossing a slow edge in the last move only and (b) $f_1(C')\neq 0$. Let's call $S_1(C)$ the set of all such configurations and $V_1(C)$ the set of all configurations that are visited in going from $C$ to all $C'\in S_1(C)$ (not including $C'$ and $C$). Starting from the equation for $f_1(C)$, $C\in B$, 

\begin{equation}
\eqalign{\mathcal{A}_\gamma(C)f_1(C)&=\sum_{C''}W(C''\rightarrow C)(1-\delta_{\gamma,W(C'\rightarrow C)})f_{2}(C'')\cr
&+\sum_{C''}\delta_{\gamma,W(C''\rightarrow C)}f_{1}(C''),\quad C\in B,\cr}
\label{master_f2a}
\end{equation}

\noindent the idea is to eliminate $W(C''\rightarrow C)f_2(C'')$ by noting that 

\begin{equation}
\eqalign{\underbrace{\mathcal{A}_0(C'')}_{=1}f_2(C'')&+\mathcal{A}_\gamma(C'')f_1(C'')=\sum_{C'''}\delta_{\gamma,W(C'''\rightarrow C'')}f_{1}(C''')\cr
&+\sum_{C'''}W(C'''\rightarrow C'')(1-\delta_{\gamma,W(C'''\rightarrow C'')})f_{2}(C'''),\cr}
\label{master_f2b}
\end{equation}

\noindent This time, however, $f_1(C'')$ on the left is not necessarily zero. To eliminate $f_2(C'')$ in (\ref{master_f2a}), we add $\mathcal{A}_\gamma(C'')f_1(C'')$ to both sides of (\ref{master_f2a}) and substitute $\mathcal{A}_0(C'')f_2(C'')+\mathcal{A}_\gamma(C'')f_1(C'')$ with (\ref{master_f2b}). By repeating the process of moving particles backwards and eliminating any $\mathcal{A}_0(C'')f_2(C)$ by adding and subtracting $\mathcal{A}_\gamma(C'') f_1(C'')$, we finally get a closed system of equations

\begin{equation}
\fl\mathcal{A}_\gamma(C)f_1(C)=\sum_{C'\in S_1(C)}f_1(C')-\sum_{C''\in V_1(C)}\mathcal{A}_{\gamma}(C'')f_1(C''), \quad C\in B.
\label{system_f1}
\end{equation}

\noindent Since $S_1(C)$ also contains blocked configurations, we can rewrite (\ref{system_f1}) as

\begin{equation}
\mathcal{A}_\gamma(C)f_1(C)-\sum_{C'\in S_1(C) \cap B}f_1(C')=h_1(C), \quad C\in B,
\label{system_f1b}
\end{equation}

\noindent where $h_1(C)$ is given by

\begin{equation}
\label{h_1}
h_1(C)=\sum_{C'\in S_1(C) \atop C'\notin B}f_1(C')-\sum_{C'\in V_1(C)}\mathcal{A}_{\gamma}(C')f_{1}(C'),\quad C\in B.
\end{equation}

\noindent If it weren't for the $h_1(C)$, (\ref{system_f1b}) would be just the same as in (\ref{system_f0a}). Because of the non-zero terms on the right, (\ref{system_f1b}) no longer describes a stochastic process as in (\ref{system_f0a}). It may also be much difficult to solve (\ref{system_f1}) than (\ref{system_f0a}) if $B$ is large.

\subsection{Higher-order terms}

In principle, the same procedure can be applied to higher order terms. Let's denote with $S_k(C)$ the set of all configurations $C'$ such that (a) $C'$ can be reached from $C$ by moving particles backwards and crossing a slow edge in the last move only (b) $f_k(C')\neq 0$. Then for $C\notin B$ we have

\begin{equation}
f_k(C)=\sum_{C'\in S_{k-1}(C)}\lambda(C,C')f_{k-1}(C')-\frac{\mathcal{A}_\gamma(C)}{\mathcal{A}_0(C)}f_{k-1}(C), \quad C\notin B
\label{f_k}
\end{equation}

\noindent where $\lambda(C,C')=0$ for $C'\notin S_{k-1}(C)$, and is given by (\ref{path_weights}) for $C'\in S_{k-1}(C)$. The additional term on r.h.s. of (\ref{f_k}) was not present for $k=1$ only due to the fact that $f_0(C)=0$ for $C\notin B$. 

Now, for $C\in B$, let's denote with $V_k(C)$ the set of all configurations that are visited in going from $C$ to all $C'\in S_k(C)$. To find $f_k(C)$, $C\in B$, we have to solve the following system of equations,

\begin{equation}
\fl\quad \mathcal{A}_\gamma(C)f_k(C)=\sum_{C'\in S_k(C)}f_k(C')-\sum_{C''\in V_k(C)}\mathcal{A}_{\gamma}(C'')f_{k}(C''), \quad C\in B.
\label{system_fk}
\end{equation}

\noindent As for $k=1$, we can rewrite (\ref{system_fk}) as

\begin{equation}
\label{system_fka}
\mathcal{A}_\gamma(C)f_k(C)-\sum_{C'\in S_k(C)\cap B}f_k(C')=h_k(C), \quad C\in B,
\end{equation}

\noindent where $h_k(C)$ is given by

\begin{equation}
h_k(C)=\sum_{C'\in S_k(C) \atop C'\notin B}f_k(C')-\sum_{C'\in V_k(C)}\mathcal{A}_{\gamma}f_{k}(C'), \quad C\in B.
\end{equation}

For the reasons evident in the following section, going beyond linear order becomes highly non-trivial for larger systems. In the rest of this paper we consider therefore only zeroth- and first-order terms in some simple disorder configurations. Insight that this approach gives us for general configurations of disorder is discussed in section \ref{further_applications}.


\section{Examples}
\label{examples}

\subsection{TASEP with a single slow site}
\subsubsection*{Periodic boundary conditions.}

The TASEP on a ring with a slow site placed at site $L$ has only one blocked configuration $C_P$ with respect to hopping rate $r<1$, the one in which all particles are immediately behind the slow site. It follows then that the equation (\ref{system_fk}) is trivially solved for any $k\geq 0$, i.e. $f_k(C_P)=\textrm{const.}\equiv 1$ for any $k\geq 0$. Using the fact that $f_0(C)=0$ for any $C\notin B=\{C_P\}$, we can write $f_0(C)$ using the delta Kronecker function, $f_0(C)=\delta_{C,C_P}$. To calculate the small-$r$ expansion of the current $j_L(r)$, we can choose $j_L=\langle r\tau_L(C)[1-\tau_1(C)]\rangle$, so that

\begin{equation}
\label{c2_oneslowsite}
j_L(r)=\frac{a_0}{b_0}r-\frac{b_1-a_1}{b_0} r^2+\Or(r^3),
\end{equation}

\noindent where $a_k$ and $b_k$ are given by

\begin{equation*}
a_k=\sum_C\tau_L(C)[1-\tau_1(C)]f_k(C), \quad b_k=\sum_C f_k(C), \quad k\geq 0.
\end{equation*}

\noindent From here it follows easily that $a_0=b_0=1$ and therefore $j_L(r)=r+\Or(r^2)$. To calculate the second-order term in $r$, we have to find $f_1(C)$, i.e. $\lambda(C,C_P)$ because of

\begin{equation*}
f_1(C)=\sum_{C'}\lambda(C,C')f_0(C')=\sum_{C'}\lambda(C,C')\delta_{C',C_P}=\lambda(C,C_P).
\end{equation*}

\begin{figure}[bht]
\centering\includegraphics[width=6cm]{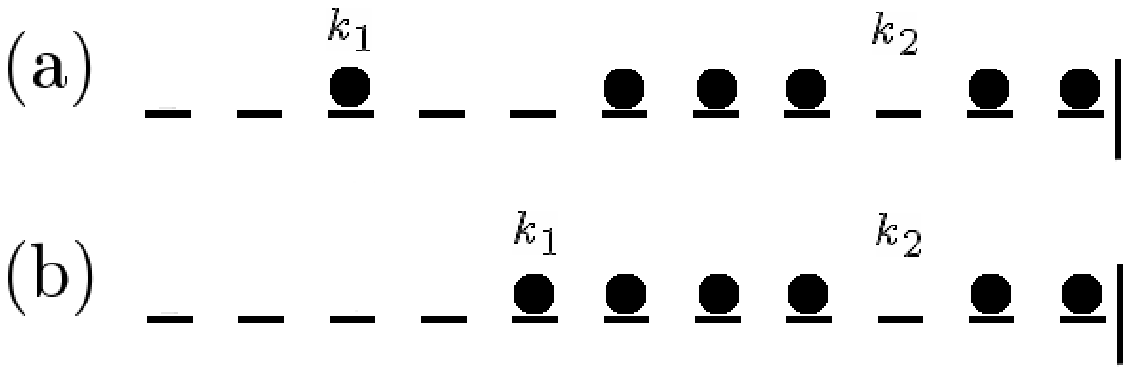}
\caption{Configurations that have non-zero first-order term $f_1(C)\neq 0$ have either (a) one particle outside the queue and one hole inside the queue or (b) one particle taken out of the queue and added to its end.}
\label{fig4}
\end{figure}

\noindent The construction of $\lambda(C,C')$, as explained in section \ref{first-order}, tells us to look for configurations $C$ such that the blocked state $C_P$ is reached by moving particles backwards from $C$, provided the slow edge is crossed only by entering $C_P$. It is easy to see that any such $C$ must have a particle displaced from the queue (figure \ref{fig4}). Thus the configurations giving non-zero first-order terms are particle-hole excitations of the blocked state $C_P$. 


\noindent To calculate $a_1$ and $b_1$, it proves useful to define a set of configurations $\mathcal{J}_L$,

\begin{equation*}
\mathcal{J}_L\equiv\{C\vert \tau_L(C)=1, \tau_1(C)=0\}.
\end{equation*}

\noindent In other words, $\mathcal{J}_L$ is simply the set of all configurations such that $r\tau_L(C)[1-\tau_L(C)]\neq 0$. Using this definition, second-order term in (\ref{c2_oneslowsite}) can be rewritten as

\begin{equation*}
c_2=-\frac{b_1-a_1}{b_0}=-\sum_{C\notin\mathcal{J}_L}\lambda(C,C_P).
\end{equation*}

\noindent This expression slightly simplifies the calculation of $c_2$, as we must calculate $\lambda(C,C_P)$ only for $C\notin\mathcal{J}_L$, and not for all $C$ having non-zero $f_1(C)$. To calculate the matrix element $\lambda(C,C_P)$ for $C\notin\mathcal{J}_L$, we can use the expression (\ref{path_weights}) derived in the previous section. All possible $C\notin\mathcal{J}_L$ having non-zero $f_1(C)$ will either have a particle at site $k_1=1,\dots,L-N-1$ and a hole at site $k_2=L$, or particles both at $k_1=1$ and $L$ with a hole at site $k_2=L-N+1,\dots,L-1$. Let's first consider  configurations $C\notin\mathcal{J}_L$ such that the particle outside the queue is placed at $k_1=1,\dots,L-N-1$ leaving a hole at site $L$. If $k_1\neq L-N-1$ (meaning that there is at least one hole in front of the particle), then $\mathcal{A}_0(C')=2$ for any $C'$ visited in going from $C$ to $C_P$. The matrix element $\lambda(C,C_P)$ for such $C$ is therefore given by

\begin{equation*}
\lambda(C,C_P)=\left(\frac{1}{2}\right)^{k_1},\quad k_1=1,\dots,L-N-1,\quad k_2=L.
\end{equation*}

\noindent If $k_1=L-N$, then $\mathcal{A}_0(C)=1$ and the rest of the configurations in going from $C$ to $C_P$ have $\mathcal{A}_0(C')=2$. This gives

\begin{equation*}
\lambda(C,C_P)=\left(\frac{1}{2}\right)^{L-N-1},\quad k_1=L-N,\quad k_2=L.
\end{equation*}

\noindent Similarly, in going from configurations with $k_1=1$ and $k_2=L-N+1,\dots,L-1$ to $C_P$, $\mathcal{A}_0$ is always $2$ and therefore $\lambda(C,C_P)$ is given by

\begin{equation*}
\lambda(C,C_P)=\left(\frac{1}{2}\right)^{L-k_2+1},\quad k_1=1,\quad k_2=L-N+1,\dots,L-1.
\end{equation*}

\noindent If $k_1=1$ and $k_2=L-N$, then $\mathcal{A}_0(C)=1$ and so $\lambda(C,C_P)$ is given by

\begin{equation*}
\lambda(C,C_P)=\left(\frac{1}{2}\right)^{N},\quad k_1=1,\quad k_2=L-N.
\end{equation*}

\noindent Summing all four contributions gives

\begin{eqnarray*}
\fl c_2&=&-\sum_{k_1=1}^{L-N-1}\left(\frac{1}{2}\right)^{k_1}-\left(\frac{1}{2}\right)^{L-N-1}-\sum_{k_2=L-N+1}^{L-1}\left(\frac{1}{2}\right)^{L-k_2+1}-\left(\frac{1}{2}\right)^{N}=\nonumber\\
\fl &=&-\left[\frac{1-\left(\frac{1}{2}\right)^{L-N}}{1-\frac{1}{2}}-1\right]-\left(\frac{1}{2}\right)^{L-N-1}-\left[\frac{1-\left(\frac{1}{2}\right)^{N+1}}{1-\frac{1}{2}}-1\right]-\left(\frac{1}{2}\right)^{N}\nonumber\\
\fl &=&-\frac{3}{2}.
\end{eqnarray*}

\noindent Our method thus gives us $j_L(r)$ up to $\Or(r^3)$

\begin{equation*}
j_L(r)=r-\frac{3}{2}r^2+\Or(r^3), \quad r\ll 1,
\end{equation*}

\noindent which was first calculated by Janowsky and Lebowitz \cite{JanowskyLebowitz94}. Unfortunately, this is as far as we can go without much effort. To calculate the next-order terms, we would have to explore paths starting from configurations having either two particles outside the queue and a hole inside the queue, or one particle outside the queue and two holes inside the queue. However, tracking movement of two particles or two holes is no longer trivial because of the exclusion, and therefore it becomes increasingly difficult, albeit possible, to calculate $\lambda(C,C')$.
   
\subsubsection*{Open boundary conditions.}

Now let's consider open boundaries case with slow site placed at site $k$. Here we can expand the current $j_L(\alpha,\beta,r)$ in any of the hopping rates $\alpha$, $\beta$ or $r$. The simplest case to consider is when none of the two remaining hopping rates are equal to the one that we are expanding in. In that case $B$ has only one configuration: an empty chain if expanding around $\alpha=0$, a fully occupied chain if expanding around $\beta=0$ and a semi-full chain with particles behind the slow site if expanding around $r=0$. When expanding $j_L$ around $r=0$, the calculation is similar to the one for the periodic case and in fact gives the same result,

\begin{equation*}
j_L(\alpha,\beta,r)=r-\frac{3}{2}r^2+\Or(r^3),\quad r\ll \alpha,\beta.
\end{equation*}

\noindent That $j_L$ does not depend on $\alpha$ nor $\beta$ in the small-$r$ limit was recognized long time ago by Janowsky and Lebowitz \cite{JanowskyLebowitz94} by studying $\alpha=\beta$ case. Note also that $j_L$ in the small $r$ limit does not depend on the position $k$ of the slow site either.

Around $\alpha=0$ or $\beta=0$, the calculation is even simpler. There is only one blocked configuration with respect to $\alpha$, and that is an empty chain. This gives us immediately $a_0=b_0=1$, i.e. $c_1=1$. The expression for $c_2$ reads

\begin{equation*}
c_2=\frac{a_1}{b_0}-c_1\frac{b_1}{b_0}=-\sum_{C\notin\mathcal{J}_0}f_1(C),
\end{equation*}

\noindent where $\mathcal{J}_0=\{C\vert \tau_1(C)=1\}$. There is only one configuration $C\in\mathcal{J}_0$ that has $f_1(C)\neq 0$, and that is the  configuration with a particle at site $1$, which has $f_1(C)=1$. A similar calculation can be made for the expansion around $\beta=0$. For the first two coefficient, the final result is thus the same as for the pure TASEP,

\begin{eqnarray*}
j_L(\alpha,\beta,r)&=&\alpha-\alpha^2+\Or(\alpha^3),\quad \alpha\ll \beta,r\\
j_L(\alpha,\beta,r)&=&\beta-\beta^2+\Or(\beta^3),\quad \beta\ll \alpha,r
\end{eqnarray*}

\noindent Again we see that the coefficients are pure numbers and do not depend on other hopping rates, nor on the position of the slow site.

When $\alpha$ or $\beta$ is equal to $r$ we immediately notice that $\vert B\vert$ (the number of elements in $B$) is greater than $1$, and therefore solving (\ref{system_f1b}) cannot be avoided. It is the same difficulty that we are going to encounter when dealing with more than one slow site in the following sections. Let's consider $\alpha=r\ll\beta$ case first. Blocked configurations can be described as having a queue behind the slow site and an empty segment in front of it. Compared to the $r\ll\alpha,\beta$ case, the queue is now no longer of size $k$ (i.e. occupying the whole segment behind the slow site), but can be of any size $0,\dots,k$ giving $\vert B\vert=k+1$. Let's denote configurations belonging to $B$ with $C_m$, where $m=0,\dots,k$ is the size of the queue behind the slow site placed at $k$. The equations for $f_0(C_m)$ are easily generated using (\ref{system_f0a}) giving

\begin{eqnarray*}
f_0(C_0)&=&f_0(C_1)\\
2f_0(C_m)&=&f_0(C_{m-1})+f_0(C(m+1)),\quad m=1,\dots,k-1\\
f_0(C_k)&=&f_0(C_{k-1}).
\end{eqnarray*}

\noindent According to our interpretation using partial exclusion process, this corresponds to having a single site with capacity $k$ which exchanges particles with two reservoirs at rates $\alpha=\beta=1$. The solution to the system above is simply $f_0(C_m)=\textrm{const.}$. If we choose $g(C)=r\tau_k[1-\tau_{k+1}]$, then $a_0=k$ and $b_0=k+1$ so that $c_1=a_0/b_0=k/(k+1)$. A much more involved calculation is required to get the second-order terms. Here we state the final result leaving the details of this calculation to \ref{appendix_b}

\begin{equation}
\label{j_ar}
j_L(\alpha,\beta,r)=\frac{k}{k+1}r-\frac{(k-1)(5k+8)}{4(k+1)^2}r^2+\Or(r^3), \quad \alpha=r\ll\beta.
\end{equation}

\begin{figure}[bht]
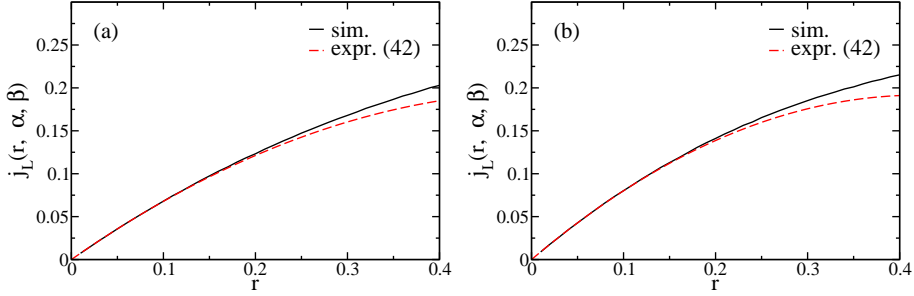

\centering\includegraphics[width=6cm]{fig5a.eps}
\centering\includegraphics[width=6cm]{fig5b.eps}
\caption{Current $j_L(r,\alpha,\beta)$ as a function of $\alpha=r$ obtained by Monte Carlo simulations (\full) on a lattice of $L=1000$ sites with $\beta=1$, compared to the expression (\ref{j_ar}) truncated at the second-order (\broken) for (a) $k=3$ and (b) $k=10$.}
\label{fig5}
\end{figure}

\noindent How well truncating the expression (\ref{j_ar}) at second-order approximates $j_L(r)$ is presented in figure \ref{fig5}, where (\ref{j_ar}) is compared to $j_L(r)$ obtained from Monte Carlo simulations on a lattice of $L=1000$ sites for $\beta=1$ and (a) $k=3$ and (b) $k=10$.


Using the particle-hole symmetry $\tau_i\leftrightarrow 1-\tau_{L-i+1}$, $\alpha\leftrightarrow\beta$, $k\leftrightarrow L-k$, we can also get the expansion for $\beta=r\ll \alpha$ which reads

\begin{equation}
\label{j_br}
\fl j_L(\alpha,\beta,r)=\frac{L-k}{L-k+1}r-\frac{(L-k-1)[5(L-k)+8]}{4(L-k+1)^2}r^2+\Or(r^3), \quad \beta=r\ll\alpha.
\end{equation}

In the most complicated case when $\alpha=\beta=r$, which corresponds to the partial exclusion process with two sites having capacities $k$ and $L-k$, unfortunately we were not able to find even $f_0(C)$, i.e. to solve (\ref{system_f0a}) for general $k$. This already clearly demonstrates that severe difficulties are to be expected whenever $\vert B\vert$ is not small.

\subsection{TASEP with two slow sites}

\subsubsection*{Periodic boundary conditions.} We next consider two slow sites placed at $k_1=L-d$ and $k_2=L$ on a ring of $L$ sites and $N$ particles. We will further assume that $d<N<L-d$\footnote{Other values of $N$ can be explored as well, but we are here mainly interested in large $N$ and small $d$.}, which means the number of particles in the segment $i=L-d+1,\dots,L$ can take values $0,\dots,d$. The number of blocked configurations is then $\vert B\vert=d+1$. The corresponding partial exclusion process consists of two sites with periodic boundary conditions, which has a simple steady state with all $f_0(C)=\textrm{const.}$. Choosing again $j_L=\langle r\tau_{k_1}(C)[1-\tau_{k_1+1}(C)]\rangle$ gives immediately $a_0=d$ and $b_0=d+1$, i.e. $c_1=d/(d+1)$, as in (\ref{j_2}). The calculation of the second-order term is very similar to the $\alpha=r$ case with a single slow site. The final result for $j_L(r)$ up to $\Or(r^3)$ is

\begin{equation}
\label{j_2_derived}
j_L(r)=\frac{d}{d+1}r-\frac{(d-1)(3d+4)}{2(d+1)^2}r^2+\Or(r^3).
\end{equation}

\begin{figure}[bht]
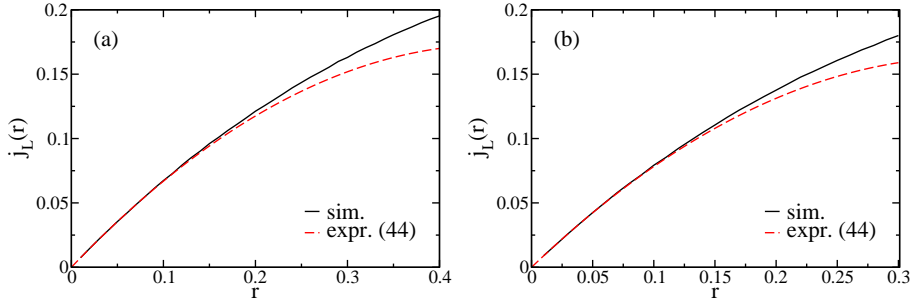

\centering\includegraphics[width=6cm]{fig6a.eps}
\centering\includegraphics[width=6cm]{fig6b.eps}
\caption{Current $j_L(r)$ as a function of $r$ obtained by Monte Carlo simulations (\full) on a ring of $L=1000$ sites and $N=500$ particles, compared to the expression (\ref{j_2_derived}) truncated at the second-order (\broken) for (a) $d=3$ and (b) $d=10$.}
\label{fig6}
\end{figure}

\noindent Notice that as $d\rightarrow\infty$, $j_\infty(r)=r-3r^2/2+\Or(r^3)$, as in the TASEP with a single slow site. A comparison of (\ref{j_2_derived}), truncated at the second-order, with $j_L(r)$ obtained from Monte Carlo simulations ($L=1000$, $N=500$) is presented in figure \ref{fig6} for (a) $d=3$ and (b) $d=10$. 


\subsubsection*{Open boundary conditions.} For $r\ll \alpha,\beta$, the result is the same as in (\ref{j_2_derived}),

\begin{equation}
\label{j_2_derived_open}
j_L(r)=\frac{d}{d+1}r-\frac{(d-1)(3d+4)}{2(d+1)^2}r^2+\Or(r^3),\quad r\ll\alpha,\beta.
\end{equation}

\noindent Other cases, i.e. when $\alpha$ or $\beta$ are equal to $r$, are more difficult to deal with as the analytical solution to (\ref{system_f0a}) is generally not known.

\subsection{TASEP with a bottleneck}

Finally, we mention the case of all slow sites being clustered in a \emph{bottleneck}, which was previously studied in \cite{ChouLakatos04}. Here the set $B$ is $2^l$\footnote{In the periodic boundaries case that is true provided $l<N$, which is the case we consider.}, where $l$ is the number of slow sites. Using the previously developed mapping to the partial exclusion process in the limit $r\rightarrow 0$, the periodic boundaries case becomes equivalent to the pure TASEP with one large but finite reservoir. If we further assume that $l<N<L-l$ so that the finite reservoir is never empty nor fully occupied, $f_0(C)$ is given by the exact solution of the pure TASEP of size $l-1$ with open boundaries and $\alpha=\beta=1$. Using the known solution of the pure TASEP with open boundaries \cite{SchutzDomany93,DEHP}, the current $j_L(r)$ up to $\Or(r^2)$ reads

\begin{equation}
j_L(r)=\frac{l+1}{4l-2}r+\Or(r^2).
\end{equation}

\noindent The same result applies to the open boundaries case for $r\ll\alpha,\beta$, which was conjectured\footnote{Although the notion that the bottleneck behaves as a small TASEP within a big one is not surprising and new, it is unclear to us whether the authors of \cite{ChouLakatos04} were actually aware that this picture is exact in the limit $r\rightarrow 0$.} in \cite{ChouLakatos04}. Unfortunately, due to large $\vert B\vert$ we were not able to find the second-order term for general $l$. (The $l=2$ case is already covered by the previous example when the distance between two slow sites is $1$.)


\section{Further applications}
\label{further_applications}

The approach developed in section \ref{general} is general and can be applied to any \emph{unidirectional} driven diffusive system in which one can identified blocked configurations with respect to one of its hopping rates. (Here the unidirectional hopping is necessary to relate $\lambda(C,C')$ to weighted backwards paths in the configuration space.) The success of this approach will mostly depend on our ability to solve (\ref{system_f0a}) and (\ref{system_f1b}). If $B=\{C_P\}$, these are trivially solved and the main problem is to find $\lambda(C,C')$. For $\vert B\vert>1$, we may try to solve (\ref{system_f0a}) analytically for small values of $\vert B\vert $ or on a computer for larger values using the analogy with the partial exclusion process.

As a further application that goes beyond binary disorder discussed so far, here we mention some results for the slow hopping rates that are not necessary all equal. This type of disorder was in mind in the original idea of MacDonald \etal \cite{MacDonaldGibbsPipkin68,MacDonaldGibbs69}, who introduced the TASEP to model the process of translation in protein biosynthesis. In a simplified description of translation, a ribosome binds to mRNA and moves along it codon by codon translating the mRNA sequence into sequence of specific amino acids. At each step, the corresponding amino acid is transported to the ribosome by tRNA. The availability (abundance) of tRNA is thus believed to be responsible for the time scale on which the ribosome moves along the mRNA. Codons with lower concentrations of corresponding tRNA will locally suppress ribosome motion across them, acting thus as slow sites. An important question, explored extensively in \cite{Zia11}, is how are protein production rates correlated with specific sequences of codons. Translated into the TASEP, the question is to determine the limiting factor for the current with respect to the strength and the positions of slow sites. Here we discuss some immediate results that stem from our approach applied to the non-binary disorder. We will not include another important ingredient for modelling translation - the fact that ribosomes bind to approximately $12$ codon sites - as it become technically difficult to do it in our approach due to the exclusion.

For two slow sites with hopping rates $r_1\neq r_2$ our approach readily gives

\begin{equation}
j_L(r)=r^{*}-\frac{3}{2}r^{*2}+\Or(r^{*3}),\quad r^*=\textrm{min}\{r_1,r_2\}\ll\alpha,\beta
\end{equation}

\noindent This result can be easily generalized to arbitrary number $M$ of slow sites provided they all have different hopping rates

\begin{eqnarray*}
j_L(r)=r^{*}-\frac{3}{2}r^{*2}+\Or(r^{*3}),&&\quad r^*=\textrm{min}\{r_i\vert i=1,\dots,M\}\ll\alpha,\beta\\
&& \quad r_i\neq r_j, \quad \textrm{$\forall i,j,\quad i\neq j$}
\end{eqnarray*}

\noindent If two slow sites however share the same hopping rate $r$, then (\ref{j_2_derived_open}) applies provided all other slow hopping rates (including $\alpha$ and $\beta$) are mutually different and not equal to $r$. What this tells us generally is that \emph{the low-current regime of the TASEP with sitewise disorder depends only on the current-minimizing subset of slow sites with equal hopping rates, regardless of other slow sites}. 

\begin{figure}[bht]
\centering\includegraphics[width=8cm]{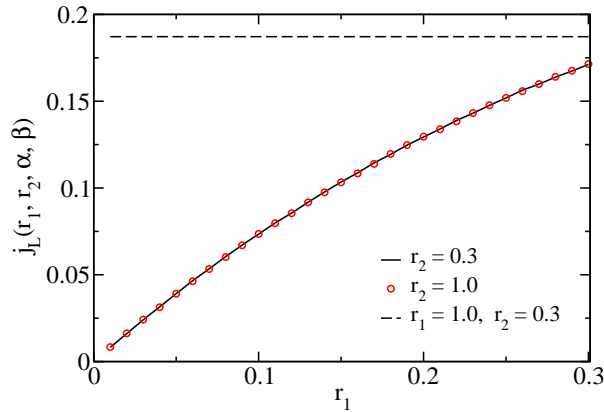}
\caption{Current $j_L(r_1,r_2,\alpha,\beta)$ obtained by Monte Carlo simulations on a lattice of $L=1000$ sites with $15$ slow sites of two types, $10$ slow sites with rates $r_1$ (type 1) and $5$ slow sites with rates $r_2$ (type 2), plotted as a function of $r_1$ for fixed $r_2=0.3$ (\full) and $r_2=1$ (\opencircle). Setting $r_2=1$ in the latter case means that only type 1 slow sites are present. Broken line (\broken) corresponds to $r_1=1$ and $r_2=0.3$ (type 2 slow sites only). In all cases $\alpha=\beta=1$. Slow sites of type $1$ were placed at $i=27,79,262,558,563,655,686,701,720,888$ and those of type $2$ at $i=156,191,407,744,997$.}
\label{fig7}
\end{figure}

Here we make a modest attempt to test this idea by simulating the TASEP with $15$ randomly distributed slow sites of which $10$ have rates $r_1$ (type 1) and $5$ have rates $r_2>r_1$ (type 2) on a lattice of $1000$ sites with $\alpha=\beta=1$. Figure \ref{fig7} compares current $j_L(r_1,r_2,\alpha,\beta)$ as a function of $r_1$ for two values of $r_2$, one with $r_2=0.3$ (both types present) and the other with $r_2=1$ (only type 1 present). Our data shows no significant difference between these two currents for small $r_1<r_2=0.3$. Notice also a gap between the current with type 2 slow sites only (\broken) and the point $r_1=r_2=0.3$ where type 1 slow sites turn into type 2, which lowers the current as there are now $15$ slow sites of type 2 instead of $5$.


While this is far from a systematic study, it shows that the low-current regime can be safely approximated by the TASEP with binary disorder provided the current-minimizing set of slow sites with equal rates is properly identified. It would be interesting to test this idea further using real data (e.g. relative abundances of tRNA measured for \emph{E. Coli} \cite{DongNilssonKurland96}) and compare it to the approximate theory of estimating currents for the disordered TASEP developed in \cite{Zia11}.


\section{Conclusion}

The ma\-trix-\-pro\-duc\-t ansatz and mean-field ap\-prox\-i\-ma\-tion are both pow\-er\-ful an\-a\-lyt\-i\-cal approaches for studying driven diffusive systems, the ASEP in particular. Unfortunately, in some cases such as site-wise disorder, it is not known how to apply the matrix-product ansatz and mean-field approximation is mostly reduced to numerical studies. In this article we showed how to access some exact steady-state properties of the TASEP with site-wise disorder in the low-current regime, i.e. when one of the hopping rates is small. 

Our approach is based on a simple fact, proved in \ref{appendix_a}, that the steady-state (non-normalized) weights are polynomials in hopping rates. Using this fact the steady-state average of any physical observable (e.g. current) can be expanded in one of the small hopping rates with coefficients that obey a specific set of equations. While our approach is not restricted particularly to the TASEP, it requires that we can identify what we call blocked configurations (configurations that the system freezes into if one of the hopping rates is set to zero). In that case the zeroth-order coefficients are non-zero only for blocked configurations, which drastically reduces the number of unknowns. For the TASEP with binary disorder where all slow sites share the same hopping rate $r<1$, we show that the zeroth-order coefficients in the small $r$ expansion are in fact steady-state weights of another (but rarely studied) process called partial exclusion process in which more than one particle per site is allowed. This mapping, which is exact in the limit $r\rightarrow 0$, is made by replacing all slow sites with boxes of capacities equal to the distances between neighbouring slow sites. A simple interpretation of this result is that the limit $r\rightarrow 0$ creates a huge separation of time scales between particles hopping at rates $r$ and $1$, so that any particle that jumps across a slow site immediately joins a queue in the front. It is remarkable (and discouraging at the same time) that the TASEP with binary disorder even in this simplified case maps to a process which itself is a hard and unsolved problem.

In cases when zeroth-order terms can still be found, we provide a recipe for calculating first-order terms, which comes down to tracking paths in the configuration space on the underlying stochastic network. In section \ref{examples} we calculated zeroth-order and in some cases first-order terms in the small $r$ expansion of the steady-state current for particular disorder configurations previously studied in \cite{JanowskyLebowitz94,ChouLakatos04}. As the number of blocked configurations increases it becomes increasingly difficult to follow this programme analytically. However, because of the mapping to the partial exclusion process the reduction of the unknowns is still huge and allows us potentially to use a computer instead, even for large lattices.

Our approach can readily be applied to non-binary disorder with more than one type of slow rates, which is relevant for modelling protein synthesis. Remarkably, the lowest-order coefficients we get by expanding in one of the slow hopping rates do not depend on the other hopping rates. In other words, the low-current regime of the TASEP with site-wise disorder depends only on the current-minimizing set of slow sites with equal hopping rates, regardless of other slow sites. Once this subset is found (which may be a hard problem in itself), one can work with binary disorder only and go from there using either the approach developed here or using e.g. phenomenological approach that looks for the largest cluster of slow sites \cite{TripathyBarma97,TripathyBarma98}.

Finally, we mention that our approach can be applied to any other driven diffusive system with particles hopping unidirectionally, provided we can identify blocked configurations with respect to one of the model's hopping rates. In the network language, unidirectional hopping is necessary to avoid loops that may exist even if one of the hopping rates is equal to zero. For example, our approach does not work for Langmuir kinetics \cite{Parmeggiani03} or a multi-lane exclusion process where particles can change lanes \cite{EvansKafriSugdenTailleur11}. However, it could be useful for studying more complex lattice geometries without resorting to mean-field approximation or even to understand why in some instances the mean-field approximation is satisfactory.
\label{conclusion}

\ack
This work has been funded in part by EPSRC under grant number EP/J007404/1. Early part of this work has been supported by the Croatian Ministry of Science, Education and Sports under grant number 035-0000000-3187.


\appendix
\section{A formal solution of the stationary master equation}
\label{appendix_a}

Consider an ergodic continuous-time Markov jump process with transition rates $W(i\rightarrow j)$. Master equation for the steady-state probabilities $P_i$ is then given by

\begin{equation}
\label{master}
0=\sum_{j}W(j\rightarrow i)P_j-\sum_{j}W(i\rightarrow j)P_j, \quad i=1,\dots,\mathcal{N},
\end{equation}

\noindent where $\mathcal{N}$ is the total number of states. Equation (\ref{master}) can be written in a more compact form by introducing a matrix $L_{ij}$,

\begin{equation}
\label{master_L}
\sum_j L_{ij}P_j=0,
\end{equation}

\begin{equation}
L_{ij}=W(j\rightarrow i)-\delta_{ij}\sum_k W(i\rightarrow k),
\end{equation}

\noindent Note that $L$ is a left stochastic matrix, which means that $\sum_i L_{ij}=0$ for any $j$. An important property of a left stochastic matrix is that it has all of its cofactors $C_{ij}=(-1)^{ij}M_{ij}$ independent of $i$. (Here $M_{ij}$ is defined to be the determinant of the submatrix of $L$ obtained by removing $i$-th row and $j$-th column from $L$.) Since we assumed that the process is ergodic, the stationary state satisfying equation $\sum_j L_{ij}P_j=0$ is unique and non-trivial, which means that $\det L=0$. By using the Laplace expansion for $\det L$ and the aforementioned fact that $C_{jk}=C_{kk}$ we get,

\begin{equation}
\det L=0=\sum_k L_{ik}C_{jk}\sum_k L_{ik}C_{kk},
\end{equation}

\noindent which means that $P_i$ must be proportional to $C_{ii}$. Since $C_{ii}$ is the determinant of a matrix whose matrix elements are linear combinations of the transition rates, we conclude that $P_i$ must be a polynomial in all the transitions rates present in (\ref{master}).


\appendix
\setcounter{section}{1}
\section{Second-order term $c_2$ for $\alpha=r$ in the TASEP with a single slow site}
\label{appendix_b}

The starting point is the expression for $c_2$

\begin{eqnarray*}
\label{alpha_r_f1}
c_2&=&\frac{a_1}{b_0}-c_1\frac{b_1}{b_0}=(a_1-b_1)\frac{c_1}{b_0}+a_1\frac{c_1-1}{b_0}\nonumber\\
&=&-\frac{k}{(k+1)^2}\sum_{C\notin\mathcal{J}_{k}}f_1(C)+\frac{1}{(k+1)^2}\sum_{C\in\mathcal{J}_k}f_1(C),
\end{eqnarray*}

\noindent where $\mathcal{J}_k=\{C\vert \tau_{k}(C)=1,\tau_{k+1}(C)=0\}$. Summations in (\ref{alpha_r_f1}) can be further separated into  

\begin{eqnarray}
\label{f1_notin_JL}
&&\sum_{C\notin\mathcal{J}_{k}}f_1(C)=\sum_{C\notin\mathcal{J}_{k} \atop C\in B}f_1(C)+\sum_{C\notin\mathcal{J}_{k} \atop C\notin B}f_1(C)\\
\label{f1_in_JL}
&&\sum_{C\in\mathcal{J}_{k}}f_1(C)=\sum_{C\in\mathcal{J}_{k} \atop C\in B}f_1(C)+\sum_{C\in\mathcal{J}_{k} \atop C\notin B}f_1(C).
\end{eqnarray}

\noindent The first sum in (\ref{f1_notin_JL}) can be greatly simplified by that fact that there is only one $C\in B$ for which $g(C)=r\tau_k(C)[1-\tau_{k+1}(C)]=0$, and that is an empty lattice,

\begin{equation*}
\sum_{C\notin\mathcal{J}_{k} \atop C\in B}f_1(C)=f_1(C_0).
\end{equation*}

\noindent To find  $f_1(C)$ for $C\in B$, we have to solve (\ref{system_f1b}), which in this case reads

\begin{equation}
\label{f1_B}
\eqalign{\fl f_1(C_0)&=f_1(C_1)+h_1(C_0)\cr
\fl 2f_1(C_m)&=f_1(C_{m-1})+f_1(C_{m+1})+h_1(C_m),\quad m=1,\dots,k-1\cr
\fl f_1(C_{k})&=f_1(C_{k-1})+h_1(C_k),\cr}
\end{equation}

\noindent where $h_1(C_m)$ is given by (\ref{h_1}). Luckily, this system admits closed expression for $f_1(C_m)$ which is

\begin{equation*}
f_1(C_m)=f_1(C_0)-\sum_{i=1}^{m}i\cdot h_1(C_{m-i}),\quad m=1,\dots,k.
\end{equation*}

\noindent The first sum in (\ref{f1_in_JL}) can be therefore written as

\begin{eqnarray*}
\sum_{C\in\mathcal{J}_{k} \atop C\in B}f_1(C)&=&k\cdot f_1(C_0)-\sum_{m=1}^{k}\sum_{i=1}^{m}i\cdot h_1(C_{m-i})\\
&=& k\cdot f_1(C_0)-\frac{1}{2}\sum_{i=0}^{k-1}(k-i)(k-i+1)h_1(C_i).
\end{eqnarray*}

\noindent Altogether, the expression for $c_2$ can be written as

\begin{equation}
\label{c_k}
\fl c_2=\frac{1}{(k+1)^2}\left[\sum_{C\in\mathcal{J}_{k} \atop C\notin B}f_1(C)-k\sum_{C\notin\mathcal{J}_{k} \atop C\notin B}f_1(C)-\frac{1}{2}\sum_{i=0}^{k-1}(k-i)(k-i+1)h_1(C_i)\right].
\end{equation}

Now, let's go back to the expression (\ref{h_1}) for $h_1(C)$. For a given $C_m$, the set $S_1(C_m)\cap B$ consists of all configurations $C$ having $f_1(C)\neq 0$ that can be reached from $C_m$ by moving particles backwards and hitting a slow edge in the last move only. Starting from $C_m$ (with $m$ particles in the segment $i=1,\dots,k$ by the definition), the resulting $C$ will have either $m-1$ or $m+1$ particles in the same segment depending on whether we crossed the left boundary or site $k$ in the last move, respectively. The resulting $C$ will necessarily have a hole at site $1$ in the former case and a particle-hole par at $k,k+1$ in the latter case. Thus by defining $\mathcal{J}_1=\{C\vert \tau_1(C)=0\}$ in the same spirit as we defined $\mathcal{J}_k$, we can write

\begin{eqnarray*}
&&\sum_{C\in S_1(C_0) \atop C\notin B}f_1(C)=\sum_{C\in\mathcal{J}_k \atop C\notin B}f_1(C)\cdot\mathbf{1}\{\sum_{i=1}^{k}\tau_i(C)=1\}\\
&&\sum_{C\in S_1(C_m) \atop C\notin B}f_1(C)=\sum_{C\in\mathcal{J}_0 \atop C\notin B}f_1(C)\cdot\mathbf{1}\{\sum_{i=1}^{k}\tau_i(C)=m-1\}\\
&&\qquad+\sum_{C\in\mathcal{J}_k \atop C\notin B}f_1(C)\cdot\mathbf{1}\{\sum_{i=1}^{k}\tau_i(C)=m+1\},\quad m=1,\dots,k-2\\
&&\sum_{C\in S_1(C_{k-1}) \atop C\notin B}f_1(C)=\sum_{C\in\mathcal{J}_0 \atop C\notin B}f_1(C)\cdot\mathbf{1}\{\sum_{i=1}^{k}\tau_i(C)=k-2\}\\
&&\sum_{C\in S_1(C_{k}) \atop C\notin B}f_1(C)=\sum_{C\in\mathcal{J}_0 \atop C\notin B}f_1(C)\cdot\mathbf{1}\{\sum_{i=1}^{k}\tau_i(C)=k-1\}
\end{eqnarray*}

\noindent where $\mathbf{1}\{X=x\}$ is an indicator function, $\mathbf{1}\{X=x\}=0$ if $X\neq $ and $1$ if $X=x$. 

Similarly, $V_1(C_m)$ in (\ref{h_1}) consists of all configurations $C$ having $f_1(C)\neq 0$ and precisely $m$ particles in the segment $i=1,\dots,k$. We can further divide this set into subsets depending on whether $C$ belongs to $\mathcal{J}_0\cap\mathcal{J}_k$ (for which $\mathcal{A}_r(C)=2$), $\mathcal{J}_0$ but not $\mathcal{J}_k$ and vice versa (for which $\mathcal{A}_r(C)=1$) or to none of these sets (for which $\mathcal{A}_r(C)=0$). The second sum in (\ref{h_1}) can be then written as

\begin{eqnarray*}
&&\sum_{C\in V_1(C_0)}f_1(C)=\sum_{C\in\mathcal{J}_0 \atop C\notin B}f_1(C)\cdot\mathbf{1}\{\sum_{i=1}^{k}\tau_i(C)=0\}\\
&&\sum_{C\in V_1(C_m)}f_1(C)=\sum_{C\in\mathcal{J}_0 \atop C\notin B}f_1(C)\cdot\mathbf{1}\{\sum_{i=1}^{k}\tau_i(C)=m\}\\
&&\qquad+\sum_{C\in\mathcal{J}_k \atop C\notin B}f_1(C)\cdot\mathbf{1}\{\sum_{i=1}^{k}\tau_i(C)=m\},\quad m=1,\dots,k-1.
\end{eqnarray*}

\noindent To ease the notation let's introduce $a_{1}^{(0)}(m)$ and $a_{1}^{(k)}(m)$ defined as

\begin{equation}
\label{a_1_0}
a_{1}^{(0)}(m)=\sum_{C\in\mathcal{J}_0 \atop C\notin B}f_1(C)\cdot\mathbf{1}\{\sum_{i=1}^{k}\tau_i(C)=m\}, \quad m=0,\dots,k,
\end{equation}

\begin{equation}
\label{a_1_k}
a_{1}^{(k)}(m)=\sum_{C\in\mathcal{J}_k \atop C\notin B}f_1(C)\cdot\mathbf{1}\{\sum_{i=1}^{k}\tau_i(C)=m\},\quad m=0,\dots,k.
\end{equation}

\noindent From here it is easy to see that 

\begin{equation*}
a_{1}^{(0)}(k)=0,\quad a_{1}^{(k)}(0)=a_{1}^{(k)}(k)=0.
\end{equation*}

\noindent We can now write $h_1(C_m)$ as

\begin{equation}
\label{h_1_a}
\eqalign{\fl h_1(C_0)&=a_{1}^{(k)}(1)-a_{1}^{(0)}(0)\cr
\fl h_1(C_m)&=a_{1}^{(k)}(m+1)-a_{1}^{(k)}(m)+a_{1}^{(0)}(m-1)-a_{1}^{(0)}(m),\quad m=1,\dots,k-2,\cr
\fl h_1(C_{k-1})&=a_{1}^{(0)}(k-2)-a_{1}^{(0)}(k-1)-a_{1}^{(k)}(k-1)\cr}
\end{equation}

\noindent Using definition for $a_{1}^{(k)}(m)$ we can also write

\begin{equation}
\label{c_k_1}
\sum_{C\in\mathcal{J}_k\atop C\notin B}f_1(C)=\sum_{m=1}^{k-1}a_{1}^{(k)}(m).
\end{equation}

\noindent While each $a_{1}^{(0)}(m)$ and $a_{1}^{(k)}(m)$ can be calculated explicitly using (\ref{path_weights}), we can further simplify calculation by showing that we must only calculate their difference $a_{1}^{(0)}(m)-a_{1}^{(k)}(m)$. Using (\ref{h_1_a}), we can show after some algebra that 

\begin{equation}
\label{c_k_3}
\eqalign{&\frac{1}{2}\sum_{i=0}^{k-1}(k-i)(k-i+1)h_1(C_i)=-k a_{1}^{(0)}(0)+\sum_{m=1}^{k-1}a_{1}^{(k)}(m)\cr
&\qquad+\sum_{m=1}^{k-1}(k-m)[a_{1}^{(k)}(m)-a_{1}^{(0)}(m)].\cr}
\end{equation}

\noindent Using (\ref{path_weights}) it is also straightforward to calculate the second sum in $c_2$. Depending on the number of particles $m$ in the segment $i=1,\dots,k$ we can show that

\begin{equation}
\label{c_k_2}
\sum_{C\notin\mathcal{J}_{k} \atop C\notin B}f_1(C)=a_{1}^{(0)}(0)+\frac{5}{2}(k-1).
\end{equation}

\noindent Inserting (\ref{c_k_3}), (\ref{c_k_1}) and (\ref{c_k_2}) in (\ref{c_k}) gives finally

\begin{eqnarray*}
\fl c_2&=&0,\quad k=1\\
\fl c_2&=&\frac{1}{(k+1)^2}\left\{-\frac{5}{2}k(k-1)+\sum_{m=1}^{k-1}(k-m)\left[a_{1}^{(0)}(m)-a_{1}^{(k)}(m)\right]\right\},\quad k\geq 2.
\end{eqnarray*}

\noindent The difference $a_{1}^{(0)}(m)-a_{1}^{(k)}(m)$ can be further written as

\begin{equation}
\label{a_diff}
\eqalign{&a_{1}^{(0)}(m)-a_{1}^{(k)}(m)=\sum_{C\in\mathcal{J}_0,C\notin\mathcal{J}_k \atop C\notin B}f_1(C)\cdot\mathbf{1}\{C\vert\sum_{i=1}^{k}\tau_i(C)=m\}\cr
&\qquad-\sum_{C\notin\mathcal{J}_0,C\in\mathcal{J}_k \atop C\notin B}f_1(C)\cdot\mathbf{1}\{C\vert\sum_{i=1}^{k}\tau_i(C)=m\}.\cr}
\end{equation}

\noindent Using (\ref{path_weights}) we can easily find that

\begin{equation*}
\fl\sum_{m=1}^{k-1}\sum_{C\in\mathcal{J}_0,C\notin\mathcal{J}_k \atop C\notin B}(k-m)f_1(C)\cdot\mathbf{1}\{C\vert\sum_{i=1}^{k}\tau_i(C)=m\}=\frac{(k-2)(7k-1)}{4}+\left(\frac{1}{2}\right)^{k-1}
\end{equation*}

\noindent To compute the second sum in (\ref{a_diff}) we must first identify configurations $C\notin\mathcal{J}_0$ and $C\in\mathcal{J}_k$ that give $f_1(C)\neq 0$. Since sites $1$ and $k$ must be occupied and site $k+1$ must be empty, we have only one option for $m=2,\dots,k-2$ and two options for $m=k-1$. For $m=2,\dots,k-2$, the only way we can reach a blocked configuration by moving particle backwards is to start from a configuration that has a queue behind the slow site and a particle at site $1$. The blocked configuration with $m-1$ particles is then reached by moving a particle at site $1$ backwards, which gives $\lambda(C,C_{m-1})=1$. We can thus write

\begin{equation}
\label{f_1_notin0_ink}
\eqalign{& \fl\sum_{m=1}^{k-1}\sum_{C\notin\mathcal{J}_0,C\in\mathcal{J}_k \atop C\notin B}f_1(C)\cdot\mathbf{1}\{C\vert\sum_{i=1}^{k}\tau_i(C)=m\}=\sum_{m=2}^{k-1}(k-m)\cdot 1\cr
\fl\quad &+\sum_{C\notin\mathcal{J}_0,C\in\mathcal{J}_k \atop C\notin B}f_1(C)\cdot\mathbf{1}\{C\vert\sum_{i=1}^{k}\tau_i(C)=k-1\}\cr}
\end{equation}

\noindent For $m=k-1$ there is a single hole in the otherwise full segment $i=1,\dots,k$, so in addition to moving backwards a particle at site $1$ we can reach a blocked configuration with $m=k$ particles by moving a particle from the segment $i=k+1,\dots,L$ or from the right reservoir. Computing $\lambda(C,C_k)$ is then straightforward but more complicated, because we must count all the ways in which a particle and a hole can both move until $C_k$ is reached. Instead of calculating each $\lambda(C,C_k)$ individually and then make the summation, we will use the following trick. Let's denote with $C_{k_1,k_2}$ configurations that have a hole placed at $k_1=2,\dots,k-1$ and a particle placed at $k_2=2,\dots,L-k+1$, where both $k_1$ and $k_2$ are  measured relative to the site $k$. Here $k_2=L-k+1$ denotes a particle in the right reservoir, i.e. a configuration with no particles in the segment $i=k+1,\dots,L$. For a given $k_1=2,\dots,k-1$ the equations for $f_1(C_{k_1,k_2})$ read

\begin{eqnarray*}
\fl 2f_1(C_{k_1,k_2})&=&f_1(C_{k_1-1,k_2})+f_1(C_{k_1,k_2-1}),\quad k_2=2,\dots,L-k-1\\
\fl (1+\beta)f_1(C_{k_1,L-k})&=&f_1(C_{k_1-1,k_2})+f_1(C_{k_1,L-k-1})\\
\fl f_1(C_{k_1,L-k+1})&=&f_1(C_{k_1-1,L-k+1})+\beta f_1(C_{k_1,L-k})
\end{eqnarray*}

\noindent Summing all the equations for a fixed $k_1=2\dots,k-1$ gives after some algebra

\begin{equation*}
\sum_{k_2=2}^{L-k+1}f_1(C_{k_1,k_2})=\sum_{k_2=2}^{L-k+1}f_1(C_{k_1-1,k_2})+f_1(k_1,1).
\end{equation*}

\noindent The advantage here is that we is no dependence on $\beta$. Solving this is as a recursion relation in $k_1$ we get

\begin{equation*}
\sum_{k_2=2}^{L-k+1}f_1(C_{k_1,k_2})=\sum_{k_2=2}^{L-k+1}f_1(C_{1,k_2})+\sum_{i=2}^{k_1}f_1(C_{i,1})
\end{equation*}

\noindent Both sums on the r.h.s. are now simple to calculate using (\ref{path_weights}) because either particle or hole is always fixed. The final result is

\begin{equation}
\label{f_1_notin0_ink2}
\sum_{k_1}^{k-1}\sum_{k_2=2}^{L-k+1}f_1(C_{k_1,k_2})=k-\frac{5}{2}+\left(\frac{1}{2}\right)^{k-1}.
\end{equation}

\noindent Inserting (\ref{f_1_notin0_ink2}) in (\ref{f_1_notin0_ink}) and then in the expression for $c_2$ we finally get

\begin{equation}
c_2=-\frac{(k-1)(5k+8)}{4(k+1)^2},\quad k\geq 1.
\end{equation}


\section*{References}
\begin{thebibliography}{99}

\bibitem{MarroDickman99} Marro J and Dickman D 1999 {\it Nonequilibrium phase transitions in lattice models} (New York: Cambridge University Press)
\bibitem{MacDonaldGibbsPipkin68} MacDonald C T \etal 1968 {\it Biopolymers} {\bf 6} 1-25
\bibitem{MacDonaldGibbs69} MacDonald C T and Gibbs J H 1969  {\it Biopolymers} {\bf 7} 707-25
\bibitem{NagelSchreckenberg92} Nagel K and Schreckenberg M 1992 {\it J. Phys. I France} 2221-9
\bibitem{SchutzDomany93} Sch\"{u}tz G M and Domany E 1993 {\it J. Stat. Phys.} {\bf 72} 277-96
\bibitem{DEHP} Derrida D, Evans M R, Hakim V and Pasquier V 1993 {\it J. Phys. A: Math. Theor.} {\bf 26} 1493
\bibitem{DerridaLebowitzSpeer01} Derrida D, Lebowitz J L and Speer E R 2001 {\it Phys. Rev. Lett.} {\bf 87} 150601
\bibitem{OnsagerMachlup53} Onsager L and Machlup S 1953 {\it Phys. Rev.} {\bf 91} 1505-12
\bibitem{Bertini09} Bertini L \etal 2009 {\it J. Stat. Phys.} {\bf 135} 857-72
\bibitem{Schadschneider10} Schadschneider A, Chowdhury D and Nishinari K 2010 {\it Stochastic Transport in Complex Systems: From Molecules to Vehicles} (Amsterdam: Elsevier)
\bibitem{Mallick96} Mallick L 1996 {\it J. Phys. A: Math. Theor.} {\bf 29} 5375
\bibitem{LeePopkovKim97} Lee H-W, Popkov V and Kim D 1997 {\it J. Phys. A: Math. Theor.} {\bf 30} 8497
\bibitem{Evans96} Evans M R 1996 {\it Europhys. Lett.} {\bf 36} 13
\bibitem{KrugFerrari96} Krug J and Ferrari P A 1996 {\it J. Phys. A: Math. Theor.} {\bf 29} L465-71
\bibitem{MallickMallickRajewsky99} K Mallick \etal 1999 {\it J. Phys. A: Math. Gen.} {\bf 32} 8399
\bibitem{EvansFerrariMallick09}  Evans M R, Ferrari P A and Mallick K 2009 {\it J. Stat. Phys.} {\bf 135} 217-239
\bibitem{LakatosChou03} Lakatos G and Chou T 2003 {\it J. Phys. A: Math. Gen.} {\bf 36} 2027 
\bibitem{ShawZiaLee03} Shaw L B \etal 2003 {\it Phys. Rev. E} {\bf 68} 021910
\bibitem{ChouLakatos04} Chou T and Lakatos G 2004 {\it Phys. Rev. Lett.} {\bf 93} 198101
\bibitem{Parmeggiani03} Parmeggiani A \etal 2003 {\it Phys. Rev. Lett.} {\bf 90} 086601
\bibitem{Reichenbach06} Reichenbach T \etal 2006 {\it Phys. Rev. Lett.} {\bf 97} 050603
\bibitem{EvansKafriSugdenTailleur11} Evans M R \etal 2011 {\it J. Stat. Mech.} P06009
\bibitem{SugdenEvans07} Sugden K E P and Evans M R 2007 {\it J. Stat. Mech.} P11013
\bibitem{BlytheEvans07} Blythe R A and Evans M R 2007 {\it J. Phys. A: Math. Theor.} {\bf 40} R333
\bibitem{EvansBlythe02} Evans M R and R A Blythe 2002 {\it Physica A} {\bf 313} 110-52
\bibitem{JanowskyLebowitz94} Janowsky S A and Lebowitz J L 1994 \textit{J. Stat. Phys.} \textbf{77} 35-51
\bibitem{TripathyBarma97} Tripathy G and Barma M 1997 {\it Phys. Rev. Lett.} {\bf 78} 3039-42
\bibitem{TripathyBarma98} Tripathy G and Barma M 1998 {\it Phys. Rev. E.} {\bf 58} 1911-26
\bibitem{Schnakenberg1976} Schnakenberg J 1976 {\it Rev. Mod. Phys.} {\bf 48} 571
\bibitem{SandowSchutz94} Sandow S and Sch\"{u}tz G M 1994 {\it Phys. Rev. E} {\bf 49} 2726-41
\bibitem{Schutz03} Sch\"{u}tz G M 2003 {\it J. Phys. A: Math. Theor.} {\bf 36} R339
\bibitem{Thompson11} Thompson A G \etal 2011 {\it J. Stat. Mech.} P02029
\bibitem{Zia11} Zia R K P \etal 2011 {\it J. Stat. Phys.} {\bf 144} 405-28
\bibitem{DongNilssonKurland96} Dong H \etal 1996 {\it J. Mol. Biol.} {\bf 206}(5) 649-63

\end{thebibliography}
\end{document}